\documentclass[preprint,showpacs,preprintnumbers,amsmath,amssymb
              ,prb,eqsecnum]{revtex4}

\usepackage{graphicx}
\usepackage{dcolumn}
\usepackage{bm}

\begin{document}

\preprint{two-mag.ver1}

\title{Two-magnon excitations in resonant inelastic x-ray scattering 
from quantum Heisenberg antiferromagnets}

\author{Tatsuya Nagao}
\affiliation{%
Faculty of Engineering, Gunma University, Kiryu, Gunma 376-8515, Japan}

\author{Jun-ichi Igarashi}%
\affiliation{%
Faculty of Science, Ibaraki University, Mito, Ibaraki 310-8512, Japan}

\date{\today}

\begin{abstract}
We study two-magnon spectra in resonant inelastic x-ray scattering (RIXS)
from the Heisenberg antiferromanets by extending the formula of Nomura and 
Igarashi (Phys. Rev. B 71, 035110 (2005)).
The core-hole potential in the intermediate state of RIXS gives rise to
a change in the exchange coupling between $3d$ electrons, leading to
an effective interaction between the core hole and spins of $3d$ electrons.
We derive a formula suitable to calculate the two-magnon RIXS intensities,
replacing the bare core-hole potential responsible to charge excitations
by this effective interaction creating two magnons in our previous formula.
It consists of two factors, one of which determines the incident-photon-energy
dependence and another is the two-magnon correlation function.
We evaluate the former factor for La$_2$CuO$_4$ in terms of the density of 
states of the $4p$ states obtained by the band calculation. We also calculate 
the two-magnon correlation function as a function of energy loss $\omega$ and
momentum transfer $\textbf{q}$ of the Heisenberg model on a square lattice, 
by summing up the ladder diagrams after transforming the magnon-magnon 
interaction into a separable form. 
The calculated spectra form a broad peak around 
$\omega=3J$ for $S=1/2$ on the magnetic Brillouin zone boundary and vanish 
at $\textbf{q}=(0,0)$ and $(\pi,\pi)$.
Such momentum dependence of the RIXS spectra could provide an excellent 
opportunity to study the dynamics in the Heisenberg model.

\end{abstract}

\maketitle

\section{\label{sect.1}Introduction}

Resonant inelastic x-ray scattering (RIXS) has recently attracted
much interest, since it provides valuable information of charge excitations 
in solids.
\cite{Kao96,Hill98,Hasan00,Kim02,Inami03,Kim04-1,Suga05} 
Unlike optical measurement, it can probe directly the momentum dependence 
of the excitations.
The $K$-edge resonance in transition-metal compounds is particularly useful,
because the corresponding x-ray wavelengths are an order of lattice spacing.
In this situation, the $1s$ core electron is prompted to an empty $4p$ state
by absorbing photon, then charge excitations are created in order to screen 
the core-hole potential, and finally the photo-excited $4p$ electron is
recombined with the core hole by emitting photon. In the end,
charge excitations are left with energy and momentum transferred from photons. 

For analyzing such spectra,
Nomura and Igarashi (NI) have developed a formalism
\cite{Nomura04,Nomura05,Igarashi06}  
by adapting the resonant Raman theory of Nozi\`eres and Abrahams.
\cite{Nozieres74} 
According to the formula NI have derived, 
the RIXS intensity is expressed in terms of 
the density-density correlation function in the equilibrium system
under the Born approximation to the core-hole potential.
Describing the electronic states on the multiband tight-binding model
within the Hartree-Fock approximation, and taking account of
the electron correlation within the random phase approximation,
Nomura and Igarashi have analyzed the RIXS spectra of undoped cuprates.
\cite{Nomura04,Nomura05,Igarashi06} 
The calculated spectra have reproduced well the experimental ones 
as a function of energy loss and the dependence on momentum.
The use of the Born approximation to the core-hole potential has been 
examined by evaluating higher-order corrections, and have partly been 
justified in spite of a strong core-hole potential.\cite{Igarashi06}
In addition, the RIXS spectra in NiO have recently been analyzed by
the same method, in an excellent agreement with the experiment.
\cite{Takahashi06} 
Therefore, the NI formula seems quite useful to analyze the RIXS spectra.
Note that most theoretical studies on the momentum dependence of
the RIXS spectra have been based on the numerical diagonalization 
method for small clusters with replacing the $4p$ band by a single level.
\cite{Tsutsui99,Okada06} By these numerical methods, 
it seems practically impossible to analyze the RIXS spectra 
in three dimensional systems.

Quite recently, Hill {\it et al.} have reported that
the RIXS intensity has been observed around the energy loss 
$300\sim 600$ meV in La$_2$CuO$_4$
with improving the instrumental resolution.\cite{Com1}
One scenario for the origin of the spectra is that the intensity
arises from two-magnon excitations.
In this paper, we examine this possibility by developing the NI formalism.
\cite{Nomura04,Nomura05,Igarashi06} 
As pointed out by van den Brink,\cite{vdBrink05}
the presence of the core-hole potential
in the intermediate state modifies the exchange process, giving rise to
the change in the exchange coupling between the spins of $3d$ 
electrons at the core-hole site and those at the neighboring sites.
This leads to an effective interaction which creates two magnons 
by the core hole. Replacing the bare core-hole potential responsible to charge
excitations by the effective interaction in the NI formula,
we immediately obtain the formula of the two-magnon RIXS spectra.
It consists of two factors, one of which gives the incident-photon-energy
dependence and another is the two-magnon correlation function.
The former factor involves the density of states (DOS) of the $4p$ states,
and is almost independent of energy loss. We evaluate it for La$_2$CuO$_4$,
using the $4p$ DOS given by the local density approximation (LDA). 
A large enhancement is predicted at the $K$ edge as a function of 
incident-photon energy; the enhancement is much larger for the polarization 
along the $c$ axis than for the polarization in the $ab$ plane. 

The two-magnon correlation function determines the dependence on 
the energy loss $\omega$ and the momentum transfer $\textbf{q}$.
We calculate it within the first order of $1/S$ ($S$ is the magnitude of spin)
at zero temperature on the antiferromagnetic 
Heisenberg model in a square lattice.
It is known that the $1/S$ expansion works well with taking account of 
the quantum fluctuation.
The linear-spin-wave theory is made up of a leading-order in the $1/S$ 
expansion,\cite{Anderson52,Kubo52} and the magnon-magnon interaction 
as well as the Oguchi correction to the magnon energy arise in the first order 
of $1/S$.\cite{Oguchi60,Harris71} Various physical quantities 
such as the spin-wave dispersion, the sublattice magnetization, 
the perpendicular susceptibility, and the spin-stiffness constant, 
have been calculated up to the second order in $1/S$ in the square lattice.
\cite{Igarashi92-1,Igarashi93,Canali92,Hamer92,Igarashi05}
In this paper, the magnon-magnon interaction is taken into account 
by summing up the ladder diagrams with transforming the interaction
into a separable form.  

It is found that the spectral shape as a function 
of energy loss is strongly modified by the interaction.
Such an effect is known in light scattering, although the momentum
transfer is limited to zero.\cite{Fleury68,Elliott69,Parkinson69,Singh89-3,
Canali92-2,Sandvik98}
RIXS could detect the momentum dependence of the spectra.
Our results show that
the spectra form a broad peak around $\omega=3J$ ($J$ is the exchange constant)
for $S=1/2$ at the magnetic Brillouin zone boundary and vanish
at $\textbf{q}=(0,0)$ and $(\pi,\pi)$.
Note that the momentum dependence of two-magnon excitations has been studied
in the context of the phonon-assisted photon absorption spectra,
\cite{Lorenzana95} and has recently been discussed in light scattering.
\cite{Donkov06} The latter case may be relevant to the 
non-resonant inelastic x-ray scattering experiments,
but the mechanism causing two-magnon excitations is quite different from
RIXS without core hole. This may make the RIXS spectra quite different
from the light scattering spectra because of the different matrix elements.

The present paper is organized as follows.
In Sec. \ref{sect.2}, we formulate the RIXS spectra for the Heisenberg 
antiferromagnets by adapting the NI formalism.
In Sec. \ref{sect.3}, the RIXS formula is expressed with magnon operators.
The two-magnon correlation function is calculated within the first order 
in the 1/S expansion. 
In Sec. \ref{sect.4}, numerical results are presented for the RIXS spectra 
in a two-dimensional Heisenberg model for La$_2$CuO$_4$.
Section \ref{sect.5} is devoted to the concluding remarks.
The spectra (at $\textbf{q}=0$) in light scattering are
summarized in Appendix for the square lattice.

\section{\label{sect.2}Two-magnon process in RIXS}

At the transition-metal $K$ edges,
the $1s$-core electron is excited to the $4p$ band in a dipole transition 
by absorbing photon. This process may be described by 
\begin{equation}
H_{x}=w\sum_{\textbf{q}\alpha}\frac{1}{\sqrt{2\omega_{\textbf{q}}}}
\sum_{j\eta\sigma}e_{\eta}^{(\alpha)}p_{j\eta\sigma}^{\dagger}
s_{j\sigma}c_{\textbf{q}\alpha}{\rm e}^{i\textbf{q}\cdot\textbf{r}_{j}}
   +{\rm H.c.},
\end{equation}
where $e_{\eta}^{(\alpha)}$ represents the $\eta$-th component
($\eta=x,y,z$) of two kinds of polarization vectors ($\alpha=1,2$)
of photon. Operator $c_{\textbf{q}\alpha}$ stands for the annihilation
operator of the photon with momentum $\textbf{q}$ and polarization $\alpha$.
Since the $1s$ state is so localized that the $1s\to4p$
transition matrix element is well approximated as a constant $w$. 
Annihilation operators $p_{j\eta\sigma}$ and $s_{j\sigma}$
are for the $4p_{\eta}$ and $1s$ electrons
with spin $\sigma$, respectively, at the transition-metal site $j$.
The Hamiltonians for the core electron and for 
the $4p$ electrons 
are given by 
\begin{eqnarray}
H_{1s} & = & \epsilon_{1s}\sum_{j\sigma}s_{j\sigma}^{\dagger}s_{j\sigma},\\
H_{4p} & = & \sum_{\textbf{k}\eta\sigma}\epsilon_{4p}^{\eta}(\textbf{k})
p_{\textbf{k}\eta\sigma}^{\dagger}p_{\textbf{k}\eta\sigma}.
\end{eqnarray}
In the intermediate state, the attractive core-hole potential works on 
the $3d$ electrons, which may be described by 
\begin{equation}
H_{1s-3d}=V\sum_{im\sigma\sigma'}d_{im\sigma}^{\dagger}d_{im\sigma}
s_{i\sigma'}^{\dagger}s_{i\sigma'}.
\end{equation}
Annihilation operator $d_{i\sigma}$ is for the $3d$ state with spin $\sigma$
at site $i$.
Finally, for describing the low energy behavior,
we assume a single band Hubbard model for $3d$ electrons,
\begin{equation}
 H = t\sum_{<i,j>} (d^{\dagger}_{i\sigma}d_{j\sigma} + {\rm H.c.})
   + U\sum_i d_{i\uparrow}^{\dagger}d_{i\downarrow}^{\dagger}
             d_{i\downarrow}d_{i\uparrow}.
\end{equation}
Here $U$ may be $4-8$ eV, smaller than $V$.
This Hubbard model may be mapped from a more precise ``d-p" model
for cuprates.

\begin{figure}
\includegraphics[width=8.0cm]{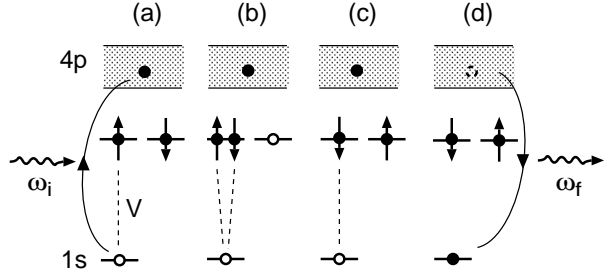}%
\caption{\label{fig.process}
A schematic representation of the two-magnon process in RIXS, corresponding 
to the first term of Eq.~(\ref{eq.jc}); (a) the $1s$ electron is excited 
to the $4p$ band by absorbing x-ray, (b,c) an exchange process takes place 
under the influence of the core-hole potential, and (d) the $4p$ electron 
is recombined with the $1s$-core hole.
}
\end{figure}

At the half-filling, a spin singlet pair has the energy $2t^2/U$ 
lower than a spin triplet pair has. Therefore, in the low energy sector,
the system may be described by the Heisenberg model with 
the exchange coupling constant $J=4t^2/U$. 
At the core-hole site, this exchange process may be influenced by
the core-hole potential, as shown in Fig.~\ref{fig.process},
which leads to the change of the exchange coupling.
This has been pointed out by van den Brink.\cite{vdBrink05}
The energy difference between the spin triplet and singlet 
of two electrons, one at the core hole site and the other
at a nearest neighbor site, is estimated as
\begin{equation}
 t^2\left[\frac{1}{U-V} + \frac{1}{U+V}\right] .
\label{eq.jc}
\end{equation}
The first and second terms arise from the process that two electrons are 
on the core-hole site as shown in Fig.~\ref{fig.process}(b) and that 
two electrons are on the nearest neighbor sites to the core-hole sites, 
respectively.
Taking the difference of the energy from the value without the core hole,
we obtain an effective interaction between the core hole at site $n$ 
and spins of $3d$ electrons,
\begin{equation}
 H_{1s-3d}^{\rm eff}=J_c \sum_{\sigma,\delta} s_{n\sigma}s^{\dagger}_{n\sigma}
                   \textbf{S}_n\cdot \textbf{S}_{n+\delta} ,
\end{equation}
with
\begin{equation}
 J_c = \frac{4t^2}{U}\frac{V^2}{U^2-V^2},
\end{equation}
where $n+\delta$ represents a nearest neighbor site to site $n$.

\begin{figure}
\includegraphics[width=8.0cm]{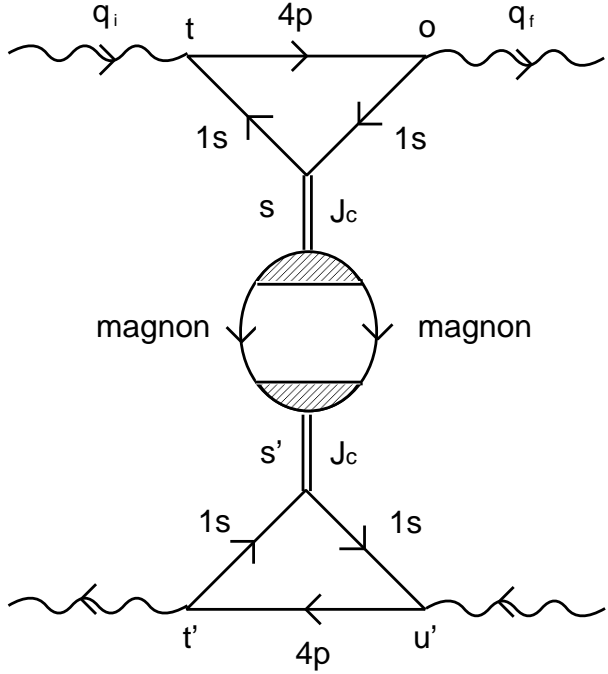}%
\caption{\label{fig.rixs}
Diagrams for the RIXS intensity in the Keldysh scheme.
The double lines represent the effective interaction between the core hole
and spins of $3d$ electrons. The solid lines with 4p and 1s
denote the Green's functions of the $4p$ electron and of 
the $1s$ core hole, respectively. The bubble with shaded vertexes
stands for the two-magnon correlation function, which connects the outward
time leg on the top half and the backward time leg on the bottom half.
}
\end{figure}

The RIXS intensity is simply given by replacing $H_{1s-3d}$ with 
$H_{1s-3d}^{\rm eff}$ in the NI formula.\cite{Nomura05}
The corresponding diagram in the Keldysh scheme is shown in 
Fig.~\ref{fig.rixs}, where the effective interaction
$H_{1s-3d}^{\rm eff}$ is represented by the double line. 
The incident photon has a momentum $\textbf{q}_i$
and an energy $\omega_i$, 
a polarization $\textbf{\textrm e}^{(\alpha_i)}$,
and the scattered photon has a momentum $\textbf{q}_f$ and an energy 
$\omega_f$, a polarization 
$\textbf{\textrm e}^{(\alpha_f)}$. The momentum and the energy
transferred from photon are given by
$\textbf{q}=\textbf{q}_i-\textbf{q}_f$ and $\omega=\omega_i-\omega_f$.
They are written simply as $q=(\textbf{q},\omega)$.
Similarly, we introduce the notations 
$q_i=(\textbf{q}_i,\omega_i)$ and $q_f=(\textbf{q}_f,\omega_f)$. 

The upper triangle represents the product of Green's functions of 
the $4p$ electron and the core hole on the outward time leg, which gives
a factor,
$\exp[i(\epsilon_{4p}^{\eta}({\textbf p})-\epsilon_{1s}-i\Gamma_{1s}-\omega_i)t]$,
with $\Gamma_{1s}$ being a lifetime broadening width of the $1s$ core hole. 
The lower triangle represents the product of Green's functions on the backward
time leg, which gives a factor 
$\exp[-i(\epsilon_{4p}^{\eta}({\textbf p})-\epsilon_{1s}+i\Gamma_{1s}-\omega_i)
(t'-u')]$. Note that extra time dependent factors,
${\rm e}^{i\omega s}$ on the outward time leg 
and ${\rm e}^{-i\omega s'}$ on the backward time leg,
arise from the interaction.
Integrating the time factor combined to the above product of Green's 
functions, with respect to $s$ and $t$ in the region of $t<s<0$, $-\infty<t<0$,
we obtain
\begin{eqnarray}
 L_B^{\eta}(\omega_i;\omega) &\equiv&
   J_c\int_{-\infty}^0{\rm d}t \frac{1}{N} \sum_{\textbf p}
   \exp[i(\epsilon_{4p}^{\eta}({\textbf p})-\epsilon_{1s}-i\Gamma_{1s}-\omega_i)t]
  \int_{t}^0 {\rm d}s\,{\rm e}^{i\omega s} \nonumber \\
     &=& - J_c\int\frac{\rho^{\eta}_{4p}(\epsilon){\rm d}\epsilon}
  {(\omega_i+\epsilon_{1s}+i\Gamma_{1s}-\epsilon)
           (\omega_i-\omega+\epsilon_{1s}+i\Gamma_{1s}-\epsilon)}.
\label{eq.born}
\end{eqnarray}
Here the summation over the momentum is replaced by the integration 
over the energy associated with  
the $4p$ DOS $\rho_{4p}^{\eta}(\epsilon)$
projected onto the $\eta$ ($=x,y,z$) symmetry.
The integration with respect to $s'$ and $t'$ on the backward time leg gives 
the term complex conjugate to Eq.~(\ref{eq.born}). 
The integration with respect to $u'$ gives the energy conservation factor, 
which guarantees that $\omega$ in Eq.~(\ref{eq.born}) is the energy loss,
$\omega=\omega_i-\omega_f$.
See Ref.~\onlinecite{Igarashi06} for the detailed derivation of this function.
Since the magnon energy is usually an order of $0.5$ eV, which is much smaller
than the energy scale of the $4p$ DOS, we can safely put $\omega=0$ 
in Eq.~(\ref{eq.born}). The bubble with shaded vertexes 
in Fig. \ref{fig.rixs} denotes 
the two-magnon correlation function $Y^{+-}(\text{q},\omega)$,
\begin{equation}
 Y^{+-}(\textbf{q},\omega) = \int_{-\infty}^{\infty}
    \langle M_{\textbf{q}}^{\dagger}(s')M_{\textbf{q}}(s)\rangle
     {\rm e}^{i\omega (s'-s)}{\rm d}(s'-s) ,
\end{equation}
with
\begin{equation}
 M_{\textbf{q}} = \sqrt{\frac{2}{N}}
       \sum_{n,\delta} \textbf{S}_{n} \cdot \textbf{S}_{n+\delta}
                             {\rm e}^{-i\textbf{q} \cdot \textbf{r}_n},
\label{eq.effective}
\end{equation}
where $\langle\cdots\rangle$ indicates the average over the ground state.
Combining all the above factors, we finally obtain the expression of the RIXS 
intensity as
\begin{equation}
 W(q_i,\alpha_i;q_f,\alpha_f) 
  = \frac{|w|^4}{4\omega_i\omega_f} \frac{N}{2}
        Y^{+-}(\textbf{q},\omega) 
        \left|\sum_{\eta}e_{\eta}^{(\alpha_f)}
             L^{\eta}_B(\omega_i;\omega)
            e_{\eta}^{(\alpha_i)} \right|^2.
\label{eq.general}
\end{equation}

\section{\label{sect.3}The 1/S expansion}

We carry out systematically the $1/S$-expansion by introducing 
the Holstein-Primakoff transformation to the spin operators.\cite{Holstein40}
Assuming two sublattices in the AF ground state, we express spin operators 
by boson operators as
\begin{eqnarray}
 S_i^z &=& S - a_i^\dagger a_i ,  
 \label{eq.boson1}\\
 S_i^+ &=& (S_i^-)^\dagger = \sqrt{2S}f_i(S)a_i ,\\
 S_j^z &=& -S + b_j^\dagger b_j ,\\
 S_j^+ &=& (S_j^-)^\dagger = \sqrt{2S}b_j^\dagger f_j(S) ,
 \label{eq.boson2}
\end{eqnarray}
where $a_i$ and $b_j$ are boson annihilation operators, and
\begin{equation}
    f_\ell (S) = \left(1 - \frac{n_\ell}{2S}\right)^{1/2}\\
               = 1 - \frac{1}{2}\frac{n_\ell}{2S} -
                  \frac{1}{8}\left(\frac{n_\ell}{2S}\right)^2 + \cdots ,
\end{equation}
with $n_\ell=a_i^\dagger a_i$ and $b_j^\dagger b_j$.
Indices $i$ and $j$ refer to sites on the "up" and "down" sublattices, 
respectively.

\subsection{Heisenberg Hamiltonian}

First we apply the $1/S$ expansion to the Heisenberg Hamiltonian described by
\begin{equation}
 H= J\sum_{<i,j>}\textbf{S}_{i} \cdot \textbf{S}_{j},
\label{eq.heis}
\end{equation}
where $\langle i,j\rangle$ indicates the sum taken over nearest neighbor
pairs.
Substituting Eqs.~(\ref{eq.boson1})-(\ref{eq.boson2}) into Eq.~(\ref{eq.heis})
we expand the Hamiltonian in powers of $1/S$ as
\begin{equation}
 H = -\frac{1}{2}JS^2Nz + H_0 + H_1 + \cdots, 
\end{equation}
where $N$ and $z$ are the number of lattice sites and that of nearest neighbor
sites, respectively. The leading term $H_0$ is expressed as
\begin{equation}
 H_0 = JS\sum_{<i,j>}( a_i^\dagger a_i + b^\dagger_{j} b_{j}
           +  a_i b_{j} + a_i^\dagger b_{j}^\dagger).
\label{eq.h0}
\end{equation}   
Rewriting the boson operators in the momentum space as
\begin{eqnarray}
 a_i &=& \left(\frac{2}{N}\right)^{1/2}\sum_{\textbf{k}} a_{\textbf{k}}
  \exp(i \textbf{k}\cdot \textbf{r}_i), \\
 b_j &=& \left(\frac{2}{N}\right)^{1/2}
       \sum_{\textbf{k}} b_{\textbf{k}} \exp(i \textbf{k}\cdot \textbf{r}_j),
\end{eqnarray}
we diagonalize $H_0$ by introducing the Bogoliubov transformation,
\begin{equation}
 a_{\textbf{k}}^\dagger = \ell_{\textbf{k}}\alpha_{\textbf{k}}^\dagger
          + m_{\textbf{k}}\beta_{-\textbf{k}}, \quad
 b_{-\textbf{k}} = m_{\textbf{k}}\alpha_{\textbf{k}}^\dagger
          + \ell_{\textbf{k}}\beta_{-\textbf{k}}, 
\label{eq.magnon}
\end{equation}
where
\begin{equation}
 \ell_{\textbf{k}} = \Bigl[\frac{1+\epsilon_{\textbf{k}}}
 {2\epsilon_{\textbf{k}}}\Bigr]^{1/2},\quad
  m_{\textbf{k}} = -\Bigl[\frac{1-\epsilon_{\textbf{k}}}
  {2\epsilon_{\textbf{k}}}\Bigr]^{1/2} \equiv - x_{\textbf{k}}\ell_{\textbf{k}},
\end{equation}
with
\begin{equation}
 \epsilon_{\textbf{k}} = \sqrt{1-\gamma_{\textbf{k}}^2}, \quad
 \gamma_{\textbf{k}} = \frac{1}{z}\sum_{\mbox{\boldmath{$\delta$}}}
     {\rm e}^{i{\textbf{k}} \cdot \mbox{\boldmath{$\delta$}} }.
\end{equation}
Here $\mbox{\boldmath{$\delta$}}$ is the nearest neighbor vectors. 
Momentum ${\textbf k}$ is defined in the first magnetic Brillouin zone (MBZ). 

As shown in Ref.~\onlinecite{Igarashi92-1}, after the Bogoliubov 
transformation, the Hamiltonian becomes 
\begin{eqnarray}
 H_0 &=& JSz\sum_{\textbf{k}}(\epsilon_{\textbf{k}}-1) 
     + JSz\sum_{\textbf{k}} \epsilon_{\textbf{k}}
   (\alpha_{\textbf{k}}^\dagger \alpha_{\textbf{k}}
   + \beta_{\textbf{k}}^\dagger\beta_{\textbf{k}}), \\
 H_1 &=& \frac{JSz}{2S} A\sum_{\textbf{k}}\epsilon_{\textbf{k}}
 (\alpha_{\textbf{k}}^\dagger \alpha_{\textbf{k}}
+ \beta_{\textbf{k}}^\dagger\beta_{\textbf{k}})
 \nonumber \\
     &+&\frac{-JSz}{2SN}\sum_{1234}\delta_{\textbf{G}}(1+2-3-4)
     \ell_1\ell_2\ell_3\ell_4  \nonumber \\
     &\times& \biggl[\alpha_1^\dagger\alpha_2^\dagger\alpha_3\alpha_4 
     B_{1234}^{(1)}+\beta_{-3}^\dagger\beta_{-4}^\dagger\beta_{-1}\beta_{-2} 
     B_{1234}^{(2)}+4\alpha_1^\dagger\beta_{-4}^\dagger\beta_{-2}\alpha_3 
     B_{1234}^{(3)}  \nonumber \\
     &+&\bigl(2\alpha_1^\dagger\beta_{-2}\alpha_3\alpha_4 B_{1234}^{(4)}
      +2\beta_{-4}^\dagger\beta_{-1}\beta_{-2}\alpha_3 B_{1234}^{(5)}
      +\alpha_1^\dagger\alpha_2^\dagger\beta_{-3}^\dagger\beta_{-4}^\dagger
     B_{1234}^{(6)} + {\rm H.c.}\bigr)\biggr],
\label{eq.intham}
\end{eqnarray}
with 
\begin{equation}
 A=\frac{2}{N}\sum_{\textbf{k}}(1-\epsilon_{\textbf{k}}) .
 \label{eq.A}
\end{equation}
The first term in Eq.~(\ref{eq.intham}) arises from setting the products 
of four boson operators into normal product forms with respect to magnon 
operators, known as the Oguchi correction.\cite{Oguchi60}
For the square lattice, $A=0.1579$.
The second term represents the scattering of magnons. 
Momenta $\textbf{k}_1$, $\textbf{k}_2$, $\textbf{k}_3$, $\cdots$ are abbreviated
as $1,2,3,\cdots$. 
The Kronecker delta $\delta_{\textbf{G}}(1+2-3-4)$ indicates
the conservation of momenta within a reciprocal lattice vector $\textbf{G}$.
Explicit expressions for $B^{(i)}$'s in a symmetric parameterization are 
given by Eqs.~(2.16)-(2.20) in Ref.~\onlinecite{Igarashi92-1}.
Here we only write down the explicit expression of $B^{(3)}_{1234}$,
which will become necessary in the next section,
\begin{eqnarray}
 B^{(3)}_{1234}&=&\gamma_{2-4}+\gamma_{1-3}x_1 x_2 x_3 x_4+\gamma_{1-4}x_1 x_2
               + \gamma_{2-3}x_3 x_4 \nonumber\\
	       &-&\frac{1}{2}(\gamma_2 x_4+\gamma_1 x_1 x_2 x_4
	        +\gamma_{2-3-4}x_3+\gamma_{1-3-4}x_1 x_2 x_3+\gamma_4 x_2 
	 \nonumber\\
	       &+&\gamma_3 x_2 x_3 x_4 + \gamma_{4-2-1}x_1 
	          +\gamma_{3-2-1}x_1 x_3 x_4) .
\end{eqnarray}

\subsection{Two-magnon operator} 

Inserting Eqs.~(\ref{eq.boson1}) $\sim$ (\ref{eq.boson2}) into 
Eq.~(\ref{eq.effective}), we expand $M_{\textbf{q}}$ in terms of boson 
operators as
\begin{eqnarray}
 M_{\textbf{q}} &=& S \sqrt{\frac{2}{N}} \sum_{\delta}
         \biggl[\sum_{i\in A}(a_i^{\dagger}a_i 
            + b_{i+\delta}^{\dagger}b_{i+\delta}+a_i b_{i+\delta}
   + a_i^{\dagger}b_{i+\delta}^{\dagger})
         {\rm e}^{i \textbf{q} \cdot \textbf{\textrm{r}}_i}
	    \nonumber\\
           & &+ \sum_{j\in B}(b_j^{\dagger}b_j 
            + a_{j+\delta}^{\dagger}a_{j+\delta}+b_j a_{j+\delta}
   + b_j^{\dagger}a_{j+\delta}^{\dagger})
         {\rm e}^{i \textbf{q} \cdot \textbf{\textrm{r}}_j }\biggr].
\label{eq.mq1}
\end{eqnarray}
Note that the momentum transfer $\textbf{q}$ is defined in the first BZ,
which is the double of the first MBZ. When $\textbf{q}$ is outside the first
MBZ, it can be brought back to the first MBZ by a reciprocal vector 
$\textbf{G}_0$, that is, $\textbf{q}=\textbf{q}_0+\textbf{G}_0$
with $\textbf{q}_0$ being inside the first MBZ. In this situation,
${\rm e}^{i \textbf{q} \cdot \textbf{\textrm{r}}_j}
=-{\rm e}^{i \textbf{q}_0 \cdot \textbf{\textrm{r}}_j}$
in the second term of Eq.~(\ref{eq.mq1}). Noting this fact and
substituting Eq.~(\ref{eq.magnon}) into Eq.~(\ref{eq.mq1}),
we obtain the expression of $M_{\textbf{q}}$ as
\begin{equation}
 M_{\textbf{q}} = \sqrt{\frac{2}{N}} \sum_\textbf{k} N(\textbf{q},\textbf{k})
   \alpha_{[\textbf{q}_0+\textbf{k}]}^{\dagger}\beta_{-\textbf{k}}^{\dagger}
   + {\textrm H. c.} + \cdots,
\end{equation}
with
\begin{eqnarray}
 N(\textbf{q},\textbf{k}) &=& Sz\bigl\{
   (1\pm\gamma_{\textbf{q}_0})\ell_{[\textbf{q}_0+\textbf{k}]}m_{\textbf{k}}
  +\textrm{sgn}(\gamma_{\textbf{G}})(\pm 1+\gamma_{\textbf{q}_0})
   m_{[\textbf{q}_0+\textbf{k}]} \ell_{\textbf{k}} \nonumber\\
 &+&[\gamma_{[\textbf{q}_0+\textbf{k}]}
   \pm \textrm{sgn}(\gamma_{\textbf{G}})\gamma_{\textbf{k}}]
    m_{[\textbf{q}_0+\textbf{k}]}m_{\textbf{k}}
  +[\gamma_{\textbf{k}}\pm \textrm{sgn}(\gamma_{\textbf{G}})
    \gamma_{[\textbf{q}_0+\textbf{k}]}]
   \ell_{[\textbf{q}_0+\textbf{k}]}\ell_{\textbf{k}}\bigr\}.
\end{eqnarray}
Sign $\pm$ corresponds to the case that $\textbf{q}$ is inside or outside 
the first MBZ.
Symbol $[\textbf{q}_0+\textbf{k}]$ stands for the $\textbf{q}_0+\textbf{k}$ 
reduced in the first MBZ
by a reciprocal lattice vector $\textbf{G}$, that is,
$[\textbf{q}_0+\textbf{k}]=\textbf{q}_0+\textbf{k}-\textbf{G}$, 
and $\textrm{sgn}(\gamma_{\textbf{G}})$ denotes the sign
of $\gamma_{\textbf{G}}$. 
This result will be used to calculate $Y^{+-}(\text{q},\omega)$ in the next 
section.  Note that $N(\textbf{q},\textbf{k})=0$ for $\textbf{q}=0$ and 
$\textbf{q}=(\pi,\pi)$ in the square lattice, indicating that no RIXS signal 
would be generated.

\subsection{Two-magnon correlation function}

Defining the two-magnon Green's function as 
\begin{equation}
F(\textbf{q}_0,\omega;\textbf{k},\textbf{k}')
=-i \int \textrm{e}^{i \omega t} {\rm d}t
\langle T[\beta_{-\textbf{k}}(t)
          \alpha_{[\textbf{q}_0+\textbf{k}]}(t)
          \alpha_{[\textbf{q}_0+\textbf{k}']}^{\dagger}
          \beta_{-\textbf{k}'}^{\dagger}] \rangle, 
\end{equation}
we rewrite the two-magnon correlation function $Y^{+-}(\text{q},\omega)$ as
\begin{equation}
Y^{+-}(\textbf{q},\omega) 
= \frac{2}{N} \sum_{\textbf{k}} \sum_{\textbf{k}'}
N(\textbf{q},\textbf{k}) N(\textbf{q},\textbf{k}')
(-2){\rm Im}F(\textbf{q}_0,\omega;\textbf{k},\textbf{k}').
\label{eq.yt}
\end{equation}
Here {\textit T} is the time-ordering operator. 
The two-magnon Green's function is expanded in terms of the one-magnon Green 
functions,
\begin{eqnarray}
 G_{\alpha\alpha}({\textbf{k}},t) &=& -i \langle T[\alpha_{\textbf{k}}(t)
 \alpha_{\textbf{k}}^\dagger (0)] \rangle, \\
 G_{\beta\beta}(\textbf{k},t) &=& -i \langle T[\beta_{{\textbf{k}}}(t)
 \beta_{\textbf{k}}^{\dagger}(0)] \rangle.
\end{eqnarray}
The unperturbed ones corresponding to $H_0$ are given by
\begin{equation}
 G_{\alpha\alpha}^{(0)}(\textbf{k},\omega)
=G_{\beta\beta}^{(0)}(\textbf{k},\omega) 
 = [\omega - \epsilon_{\textbf{k}} + i\eta]^{-1}, \quad \eta\to 0^{+},
\end{equation}
in the energy unit of $JSz$. Hereafter we use this energy unit.

In the lowest order, the two-magnon Green's function is simply given by
\begin{equation}
F(\textbf{q}_0,\omega;\textbf{k},\textbf{k}')
         =F_0(\textbf{q}_0,\omega;\textbf{k})\delta_{\textbf{k},\textbf{k}'},
\end{equation}
with 
\begin{eqnarray}
  F_0(\textbf{q}_0,\omega;\textbf{k}) 
   &=& i\int G^{(0)}_{\alpha\alpha}([\textbf{q}_0+\textbf{k}],\omega+k_0)
             G^{(0)}_{\beta\beta}(-\textbf{k},-k_0) \frac{d k_0}{2 \pi}
     \nonumber\\
   &=& \frac{1}{\omega-\epsilon_{[\textbf{q}_0+\textbf{k}]}
                      -\epsilon_{\textbf{k}}+i\eta}.
\label{eq.f0}
\end{eqnarray}
Inserting this relation into Eq.~(\ref{eq.yt}),
we obtain the correlation function in the lowest order as
\begin{equation}
 Y^{(0)+-}(\textbf{q},\omega)=2\pi \frac{2}{N}
  \sum_{\textbf{k}}N(\textbf{q},\textbf{k})^2
                        \delta(\omega-\epsilon_{[\textbf{q}_0+\textbf{k}]}
                                     -\epsilon_{\textbf{k}}).
\end{equation}

\begin{figure}
\includegraphics[width=8.0cm]{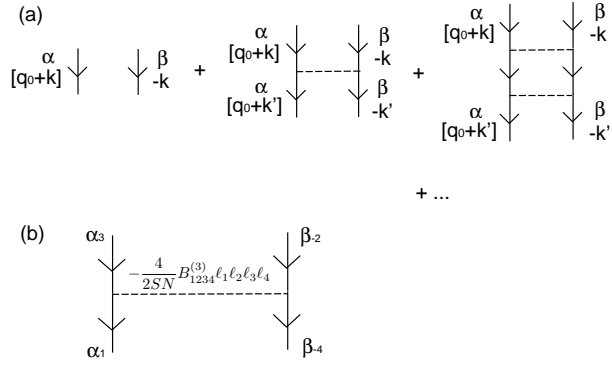}%
\caption{\label{fig.ladder}
(a) Ladder diagrams for $F(\textbf{q}_0,\omega;\textbf{k},\textbf{k}')$.
Solid lines represent the single-magnon Green's function with the magnon 
energy including the Oguchi correction.
(b) Magnon-magnon interaction in units of $JSz$.
}
\end{figure}

In the first order in $1/S$, the magnon energy $\epsilon_{\textbf{k}}$ is 
changed
into $(1+\frac{A}{2S})\epsilon_{\textbf{k}}$ due to the Oguchi correction. 
Therefore, $F_0(\textbf{q},\omega;\textbf{k})$ is modified by replacing
$\epsilon_{[\textbf{q}_0+\textbf{k}]}$ and 
$\epsilon_{[\textbf{k}]}$ in Eq. (\ref{eq.f0}) with $(1+\frac{A}{2S})
\epsilon_{[\textbf{q}_0+\textbf{k}]}$ 
and $(1+\frac{A}{2S})\epsilon_{\textbf{k}}$, respectively. 
Let $\bar{F}_0(\textbf{q},\omega;\textbf{k})$ be the function after this
modification.
Within the same order, we have to take account of the magnon-magnon 
interaction. Only the terms of a factor $B^{(3)}_{1234}$ 
in Eq.~(\ref{eq.intham}) is relevant to the present calculation.
Consider the ladder approximation shown in Fig.~\ref{fig.ladder}
for the two-magnon Green's function.
Regarding the dependence on $\textbf{k}$ and $\textbf{k}'$ in 
$F(\textbf{q}_0,\omega;\textbf{k},\textbf{k}')$ as a matrix 
$\hat{F}(\textbf{q}_0,\omega)$ with $\frac{N}{2}\times\frac{N}{2}$ dimensions, 
we notice that the sum of the ladder diagrams is equivalent to the expression,
\begin{equation}
 \left[\hat{F}(\textbf{q}_0,\omega)^{-1}\right]_{\textbf{k},\textbf{k}'} 
 = \bar{F}_0(\textbf{q}_0,\omega;\textbf{k})^{-1}
    \delta_{\textbf{k},\textbf{k}'} 
   +\frac{4}{2SN}\ell_{[\textbf{q}_0+\textbf{k}']}\ell_{\textbf{k}}
   \ell_{[\textbf{q}_0+\textbf{k}]} 
    \ell_{\textbf{k}'}
  B^{(3)}_{[\textbf{q}_0+\textbf{k}'],{\textbf{k},[\textbf{q}_0+\textbf{k}],
  \textbf{k}'}}.
\label{eq.scatter_formal}
\end{equation}
This is nothing but the eigenvalue equation for two-magnon excitations,
indicating that the ladder approximation together with 
the Oguchi correction to the single-magnon energy constitute 
the first-order correction in the $1/S$ expansion.

Equation (\ref{eq.scatter_formal}) is not useful for the actual calculation
of the two-magnon Green's function, because the matrix with 
$\frac{N}{2}\times\frac{N}{2}$ dimensions has to be inverted in order to
get $\hat{F}(\textbf{q}_0,\omega)$. We sum up exactly the ladder diagrams,
transforming the interaction into a separable form with several channels,
\begin{equation}
 -\frac{4}{2SN}\ell_1\ell_2\ell_3\ell_4 B_{1234}^{(3)} 
   = \sum_{m,n=1}^{N_c} v_{m}(2,3) \Gamma_{m n} v_{n}(4,1).
\end{equation}
Here $N_c$ is the channel number. The indices $2$ and $3$ specify the incoming 
magnons while $1$ and $4$ specify the outgoing magnons 
(Fig. \ref{fig.ladder}(b)).
Explicit forms of $v_n(\textbf{k},\textbf{k}')$ and $\Gamma_{mn}$ are 
given for the square lattice in the next section.
Thereby we obtain the T-matrix $\Pi$,
\begin{equation}
 \Pi(\textbf{q}_0,\omega:\textbf{k},\textbf{k}')
=
\frac{2}{N} \sum_{m,n} 
v_{m}(\textbf{k},[\textbf{q}_0+\textbf{k}])
\Gamma^{\rm eff}_{mn}(\textbf{q}_0,\omega)
v_{n}(\textbf{k}',[\textbf{q}_0+\textbf{k}']), 
\end{equation} 
where
\begin{equation}
\Gamma^{\rm eff}_{mn}(\textbf{q}_0,\omega)
=\left[[\textbf{I}-\hat{\Gamma}\hat{R}(\textbf{q}_0,\omega)]^{-1}
\hat{\Gamma}\right]_{mn},
\label{eq.Gmatrix}
\end{equation}
with
\begin{equation}
[{\hat R}(\textbf{q}_0,\omega)]_{mn} = \frac{2}{N} \sum_{\textbf{k}} 
 v_{m}(\textbf{k},[\textbf{q}_0+\textbf{k}])
\bar{F}_0(\textbf{q}_0,\omega;\textbf{k}) 
 v_{n}(\textbf{k},[\textbf{q}_0+\textbf{k}]).
\label{eq.R}
\end{equation}
In Eq.~(\ref{eq.Gmatrix}), the unit matrix $\hat{\textbf{1}}$ 
and $\hat{R}(\textbf{q}_0,\omega)$ are in the $N_c \times N_c$ dimensions.
We calculate the two-magnon Green's function from the T-matrix,
\begin{equation}
F(\textbf{q}_0,\omega;\textbf{k},\textbf{k}')
= 
\bar{F}_0(\textbf{q}_0,\omega;\textbf{k})\left[
\delta_{\textbf{k},\textbf{k}'} + 
 \Pi(\textbf{q}_0,\omega:\textbf{k},\textbf{k}')
\bar{F}_0(\textbf{q}_0,\omega;\textbf{k}') \right].
\end{equation}
Inserting this equation into Eq.~(\ref{eq.yt}), we obtain 
$Y^{+-}(\textbf{q},\omega)$, which gives the RIXS intensity
as a function of momentum and energy transferred from photon.

\section{\label{sect.4} Calculated Results}

We apply the formulas in the preceding sections to La$_2$CuO$_4$,
which seems to be a typical two-dimensional Heisenberg antiferromagnet.
The RIXS intensity in the two-magnon region relative to that in the
charge excitation region is roughly estimated as
$(J_c/V)^2$. For La$_2$CuO$_4$, we have $(J_c/V)^2 < 0.01$,
because $J_c\sim 0.2-0.5$ eV and $V\sim 5-10$ eV.

\subsection{Incident-photon-energy dependence}

\begin{figure}
\includegraphics[width=8.0cm]{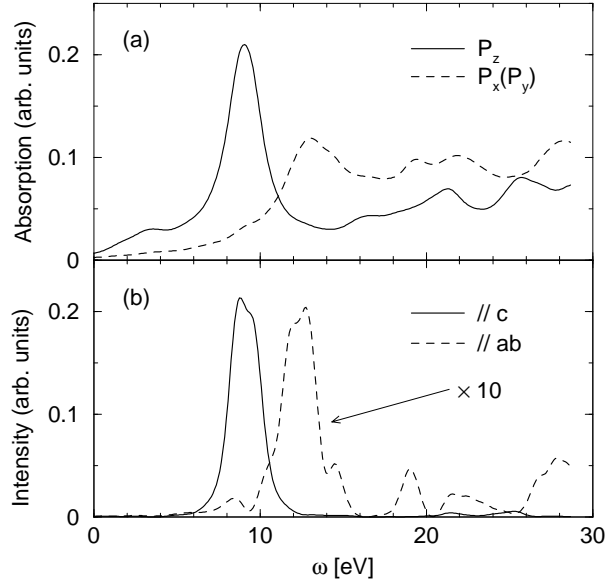}%
\caption{\label{fig.incident}
(a) The $4p$ DOS projected onto the $p_z$ and $p_x$ ($p_y$) symmetries,
calculated by the LDA in La$2$CuO$_4$.
The origin of the energy is the bottom of the $4p$ band.
(b) The RIXS intensity proportional to 
$|\sum_{\eta}e_{\eta}^{(\alpha)}
L^{\eta}_B(\omega_i;\omega=0) e_{\eta}^{(\alpha)}|^2$, 
with ${\rm e}^{(\alpha)}$ along the $c$ axis and in the $ab$ plane.
The origin of the energy corresponds to the photon energy of 
exciting the $1s$ core electron into the bottom of the $4p$ band.
}
\end{figure}

We carry out the band calculation for La$_2$CuO$_4$ within the LDA.
Figure \ref{fig.incident}(a) shows the $4p$ DOS projected onto the $p_z$
and $p_x$ ($p_y$) symmetries, which may correspond to the absorption 
coefficients for the photon polarization along the $c$ axis and in the $ab$ 
plane, respectively. 

Using the same DOS, we calculate the factor determining the 
incident-photon-energy dependence,
$|\sum_{\eta}e_{\eta}^{(\alpha)} 
L^{\eta}_B(\omega_i;\omega=0) e_{\eta}^{(\alpha)}|^2$,
for the incident and scattered photons having the same polarization.
Figure \ref{fig.incident}(b) shows the calculated results for the polarization
along the $c$-axis and in the $ab$-plane.
A strong resonant enhancement is predicted as a function of incident photon
energy. The enhancement for polarization along the $c$ axis is much stronger 
than that in the $ab$ plane.
This is quite different from light scattering where the polarization
of light is restricted in the $ab$ plane (see Appendix).\cite{Parkinson69}

\subsection{Momentum dependence as a function of energy loss}

We consider the Heisenberg model on a square lattice. 
We put $\gamma_{\textbf{k}}=(\cos k_x+\cos k_y)/2$ with $k_x$, $k_y$ in units of
1/(lattice constant).
For the magnon-magnon interaction, we have the channel number $N_c=13$,
and $\hat{\Gamma}$ is given by
\begin{equation}
\hat{\Gamma} \equiv - \frac{4}{2SN} \left(
\begin{array}{ccccccccccccc}
 0 &-1 &-1 & 2 & 2 &-1 &-1 & 
 0 & 0 & 0 & 0 & 0 & 0 \\
-1 & 2 & 0 &-1 & 0 & 0 & 0 &
 0 & 0 & 0 & 0 & 0 & 0 \\
-1 & 0 & 2 & 0 &-1 & 0 & 0 &
 0 & 0 & 0 & 0 & 0 & 0 \\
 2 &-1 & 0 & 0 & 0 &-1 & 0 &
 0 & 0 & 0 & 0 & 0 & 0 \\
 2 & 0 &-1 & 0 & 0 & 0 &-1 &
 0 & 0 & 0 & 0 & 0 & 0 \\
-1 & 0 & 0 &-1 & 0 & 2 & 0 &
 0 & 0 & 0 & 0 & 0 & 0 \\
-1 & 0 & 0 & 0 &-1 & 0 & 2 &
 0 & 0 & 0 & 0 & 0 & 0 \\
 0 & 0 & 0 & 0 & 0 & 0 & 0 &
 2 & 0 &-1 & 0 & 0 & 0 \\
 0 & 0 & 0 & 0 & 0 & 0 & 0 &
 0 & 2 & 0 &-1 & 0 & 0 \\
 0 & 0 & 0 & 0 & 0 & 0 & 0 &
-1 & 0 & 0 & 0 & 1 & 0 \\
 0 & 0 & 0 & 0 & 0 & 0 & 0 &
 0 &-1 & 0 & 0 & 0 & 1 \\
 0 & 0 & 0 & 0 & 0 & 0 & 0 &
 0 & 0 & 1 & 0 & 2 & 0 \\
 0 & 0 & 0 & 0 & 0 & 0 & 0 &
 0 & 0 & 0 & 1 & 0 & 2 \\
\end{array}
\right).
\end{equation}
The $v_n(\textbf{k},\textbf{k}')$'s are defined in Table \ref{table.1}.
In the actual calculation, we divide the first MBZ into $512\times 512$ 
meshes,
and carry out the sum over $\textbf{k}$ in Eq.~(\ref{eq.R}) to evaluate
$\hat{R}(\textbf{q}_0,\omega)$.

\begin{table}
\caption{\label{table.1}
Definition of the coefficients $v_n(\textbf{k},\textbf{k}')$
}
\begin{ruledtabular}
\begin{tabular}{cl}
$n$ &  $v_{n}(\textbf{k},\textbf{k}')$ \\
 1 & $\frac{1}{2}\ell_{\textbf{k}} \ell_{\textbf{k}'} x_{\textbf{k}}$ \\
 2 & $\frac{1}{2}\ell_{\textbf{k}} \ell_{\textbf{k}'} \cos k_x$ \\
 3 & $\frac{1}{2}\ell_{\textbf{k}} \ell_{\textbf{k}'} \cos k_y$  \\
 4 & $\frac{1}{2}\ell_{\textbf{k}} \ell_{\textbf{k}'} 
           x_{\textbf{k}'} \cos (k_x -k_x')$ \\
 5 & $\frac{1}{2}\ell_{\textbf{k}} \ell_{\textbf{k}'} 
           x_{\textbf{k}'} \cos (k_y -k_y')$ \\
 6 & $\frac{1}{2}\ell_{\textbf{k}} \ell_{\textbf{k}'} 
           x_{\textbf{k}} x_{\textbf{k}'} \cos k_x'$ \\
 7 & $\frac{1}{2}\ell_{\textbf{k}} \ell_{\textbf{k}'} 
           x_{\textbf{k}} x_{\textbf{k}'} \cos k_y'$ \\
 8 & $\frac{1}{2}\ell_{\textbf{k}} \ell_{\textbf{k}'} \sin k_x$ \\
 9 & $\frac{1}{2}\ell_{\textbf{k}} \ell_{\textbf{k}'} \sin k_y$  \\
10 & $\frac{1}{2}\ell_{\textbf{k}} \ell_{\textbf{k}'} 
           x_{\textbf{k}'} \sin (k_x -k_x')$ \\
11 & $\frac{1}{2}\ell_{\textbf{k}} \ell_{\textbf{k}'} 
           x_{\textbf{k}'} \sin (k_y -k_y')$ \\
12 & $\frac{1}{2}\ell_{\textbf{k}} \ell_{\textbf{k}'} 
           x_{\textbf{k}} x_{\textbf{k}'} \sin k_x'$ \\
13 & $\frac{1}{2}\ell_{\textbf{k}} \ell_{\textbf{k}'} 
           x_{\textbf{k}} x_{\textbf{k}'} \sin k_y'$ \\
\end{tabular}
\end{ruledtabular}
\end{table}

\begin{figure}
\includegraphics[width=8.0cm]{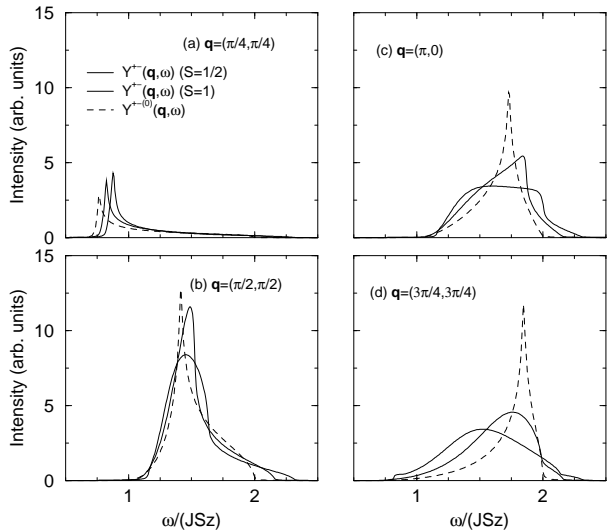}%
\caption{\label{fig.correction}
The RIXS intensity $Y^{+-}(\textbf{q},\omega)$
scaled by $(Sz)^2$ as a function of $\omega/JSz$ 
for typical values of $\textbf{q}$.
The bold and thin solid lines stand for 
$Y^{+-}(\textbf{q},\omega)/(Sz)^2$ with $S=\frac{1}{2}$ and $1$, respectively.
The broken lines represent $Y^{(0)+-}(\textbf{q},\omega)/(Sz)^2$,
which correspond to $S\to\infty$.
}
\end{figure}

Figure \ref{fig.correction} shows the calculated RIXS intensity
$Y^{+-}(\textbf{q},\omega)$ 
scaled by $(Sz)^2$ as a function of energy loss 
$\omega$ for several typical values of $\textbf{q}$.
The lowest order value, $Y^{(0)+-}(\textbf{q},\omega)/(Sz)^2$, 
is independent of $S$, corresponding to $S=\infty$. 
The effect of the magnon-magnon interaction becomes weaker with increasing $S$,
and the spectra approach to the lowest-order values.
As already pointed out, no RIXS intensity comes out at the $\Gamma$ point 
and at $\textbf{q}=(\pi,\pi)$. 
When $\textbf{q}$ is close to the $\Gamma$ point (panel (a)), 
a sharp peak is found for $\omega<JSz$, which is slightly modified in the
presence of the magnon-magnon interaction. 
When $\textbf{q}$ is at the boundary of the first
MBZ (panels (b) and (c)), a sharp peak found in the lowest order approximation 
is smeared out to be a broad peak due to the magnon-magnon interaction. 
The center of the shape remains nearly the same after taking account of 
the interaction. 
In contrast to these cases, when $\textbf{q}$ is outside the first MBZ 
(panel (d)), 
a sharp peak found in the lowest order approximation is changed into 
a broad peak with its center considerably shifted 
to the lower energy region due to the magnon-magnon interaction.

\begin{figure}
\includegraphics[width=8.0cm]{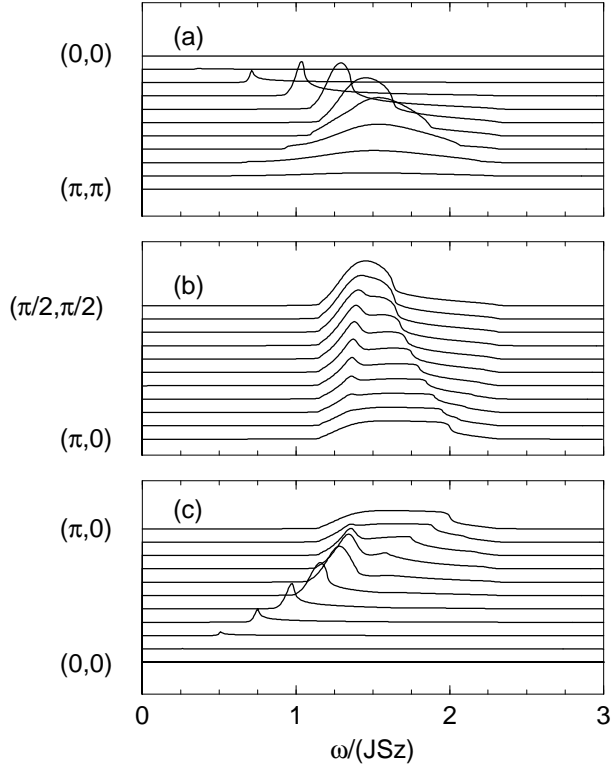}%
\caption{\label{fig.dispersion}
The RIXS intensity $Y^{+-}(\textbf{q},\omega)/(Sz)^2$ along the symmetry lines
for $\textbf{q}$ within the first order of $1/S$ ($S=1/2$).
}
\end{figure}

Figure \ref{fig.dispersion} shows the RIXS spectra as a function of
energy loss with changing momenta along several symmetry lines ($S=1/2$).
Along the zone boundary of the first MBZ (panel (b)), the spectra have large
widths around $\omega=3J$. The spectra obtained here seem to be different
from the results of van den Brink,\cite{vdBrink05} who made the moment analysis.
We find no long-lived virtual bound state of two magnons,
which has been predicted on the phonon assisted light absorption 
spectra.\cite{Lorenzana95}
Note that the matrix elements of creating two magnons in RIXS 
are different from those in light scattering.\cite{Donkov06}
Since the spectral shape depends strongly on the matrix elements,
the direct comparison of the two spectra may be less useful.
The present formalism recovers the results of light scattering
(at $\textbf{q}=0$) by Canali and Girvin,\cite{Canali92-2} as shown in Appendix.

The exchange coupling constant $J$ in La$_2$CuO$_4$ is estimated as
$J=135$ meV by comparing the spin-wave velocity calculated in the first 
order of $1/S$ with the experiment.\cite{Aeppli89}
Therefore, the broad peaks are located around $\omega=400$ meV
for $\textbf{q}$ at the boundary of the first MBZ, 
Unfortunately, no experimental data 
are available on such a low energy region in La$_2$CuO$_4$.

\section{\label{sect.5} Concluding Remarks}

We have formulated the two-magnon spectra of RIXS in antiferromagnets 
by developing the formalism of Nomura and Igarashi. The $1s$ core hole 
potential causes a change in the exchange coupling between $3d$ electrons, 
resulting in the two-magnon excitation.
This is analogous to the conventional RIXS process that the charge excitation
is created due to the screening of the core hole potential.
The intensity of two-magnon RIXS is estimated to be lass than 0.01 of
the intensity coming from the charge excitation. 

In the present formalism, the factor describing the incident-photon-energy 
dependence is separated from the factor describing the dependence on 
the momentum and the energy transferred from photon. 
We have calculated the former factor using the $4p$ DOS
for La$_2$CuO$_4$ within the LDA. We have predicted a strong enhancement of
the intensity at the $K$ edge for the polarization along the $c$ axis. 
The latter factor is given by the two-magnon correlation function. 
We have calculated the correlation function up to the first order of $1/S$ 
in the square lattice, systematically applying the $1/S$ expansion.
We have exactly summed up the ladder diagrams by transforming
the magnon-magnon interaction into a separable form with 13 channels.
The spectral shape as a function of energy loss is strongly modified 
by the magnon-magnon interaction.
On the boundary of the first MBZ, for example, the sharp peaks found in the
lowest-order approximation have considerably been broadened. We hope
the spectra obtained in this paper would be compared with the experimental
data in future.

No AF-long-range order could exist at finite temperatures
in purely two-dimensional Heisenberg models.
In the absence of the AF order, however, 
it is known from the non-linear $\sigma$ model analysis\cite{Chakravarty89}
that the spin-spin correlation length is rather long up to $T\sim J/k_B$ 
The spin-wave-like excitations could exist in such a situation. 
\cite{Tyc89,Makivic91,Nagao98}
Therefore, the present analysis of the RIXS spectra at zero temperature may
have relevance to the spectra at finite temperatures. 
The analysis of temperature effects is left in future study.

\begin{acknowledgments}
We would like to thank T. Nomura for valuable discussions.
This work was partially supported by a Grant-in-Aid for Scientific Research 
from the Ministry of Education, Culture, Sports, Science and Technology
of the Japanese Government.

\end{acknowledgments}

\appendix
\section{Two-magnon light scattering}
We summarize the two-magnon excitation of light scattering to compare
the result by the present formalism with previous studies 
in two dimensions.

Since the wavelength of light is much longer than the lattice spacing,
the intensity is independent of the directions of the incident and scattered 
lights. The interaction of light with spins is described by\cite{Parkinson69}
\begin{equation}
 H_{R} = \sum_{j \mbox{\boldmath{$\delta$}}}\left[
  A(\textbf{e}_i \cdot \mbox{\boldmath{$\delta$}})
   (\textbf{e}_f \cdot \mbox{\boldmath{$\delta$}})
 +B(\textbf{e}_i \cdot \textbf{e}_f)+C e_i^z e_f^z\right]
  (\textbf{S}_j\cdot\textbf{S}_{j+ \mbox{\boldmath{$\delta$}}}),
\end{equation}
where $\mbox{\boldmath{$\delta$}}$ 
is a unit vector in the direction joining the nearest
neighbor pairs and $A, B$, and $C$ are real constants.
The terms with $B$ and $C$ cannot cause scattering, since they
are proportional to $\sum_{j \mbox{\boldmath{$\delta$}}}\textbf{S}_j\cdot
\textbf{S}_{j+ \mbox{\boldmath{$\delta$}}}$
and commute with the magnetic Hamiltonian. 
For the polarization picking up the $A_{1g}$ mode,
$\textbf{e}_i=(\hat{\textbf{x}}+\hat{\textbf{y}})/\sqrt{2}$,
$\textbf{e}_f=(\hat{\textbf{x}}+\hat{\textbf{y}})/\sqrt{2}$,
no intensity comes out by the same reason 
($\hat{\textbf{x}}$ and $\hat{\textbf{y}}$ are unit vectors pointing 
to the $x$ and $y$ axes).
For the polarization picking up the $B_{1g}$ mode,
$\textbf{e}_i=(\hat{\textbf{x}}+\hat{\textbf{y}})/\sqrt{2}$,
$\textbf{e}_f=(\hat{\textbf{x}}-\hat{\textbf{y}})/\sqrt{2}$,
we have
\begin{equation}
 H_{R} = A\sum_{j}\left(\textbf{S}_j\cdot \textbf{S}_{j+\hat{\textbf{x}}}
                    -\textbf{S}_j\cdot \textbf{S}_{j+\hat{\textbf{y}}}\right).
\end{equation}
Expanding this in terms of magnon operators, we obtain
\begin{equation}
 H_{R} = ASz\sum_{\textbf{k}}\frac{\gamma^{d}_{\textbf{k}}}
 {\epsilon_{\textbf{k}}}
  \left( \alpha_{\textbf{k}}^{\dagger}\beta_{-\textbf{k}} + {\rm H.c.} + \cdots
	  \right),
\end{equation}
where $\gamma^d_{\textbf{k}}=(\cos k_x - \cos k_y)/2$.
The scattering intensity $I(\omega)$ from this interaction is given by
\begin{equation}
 I(\omega)\propto (ASz)^2\sum_{\textbf{k},\textbf{k}'}\left(
 \frac{\gamma^{d}_{\textbf{k}}\gamma^{d}_{\textbf{k}'}}{\epsilon_{\textbf{k}}
 \epsilon_{\textbf{k}'}}
 \right)
 (-2){\rm Im}F(\textbf{q}=0,\omega;\textbf{k},\textbf{k}').
\end{equation}
We calculate $F(\textbf{q}=0,\omega;\textbf{k},\textbf{k}')$
by summing up the ladder diagrams including the Oguchi correction
within the present formalism. We obtain the spectrum identical to that of
Canali and Girvin, which is formed by a single peak at $\omega=3.38$ for
$S=1/2$.\cite{Canali92-2}


\bibliography{mag1}

\end{document}